\documentclass[prl,aps,superscriptaddress,twocolumn,letter,nopacs,footinbib,notitlepage]{revtex4-1}

\usepackage{graphicx,color,amsmath,amsfonts,enumerate,amsthm,amssymb,mathtools,enumitem,thmtools,hyperref,subfigure,mathdots,enumitem,centernot,bm,soul,bbm,tikz,pgfplots}
\usepackage[capitalise, noabbrev]{cleveref}
\hypersetup{colorlinks=true,linkcolor=blue,citecolor=blue,urlcolor=blue}

\def\>{\rangle}
\def\<{\langle}

\newcommand{\ket}[1]{| {#1} \rangle}
 
\newcommand{\ketbra}[2]{\ensuremath{\left|#1\right\rangle\!\!\left\langle#2\right|}}
\newcommand{\braket}[2]{\ensuremath{\!\!\left\langle#1|#2\right\rangle}\!\!}
\newcommand{\matrixel}[3]{\ensuremath{\left\langle #1 \vphantom{#2#3} \right| #2 \left| #3 \vphantom{#1#2} \right\rangle}}

\renewcommand{\v}[1]{\ensuremath{\boldsymbol #1}}

\newlength{\figheight}
\newlength{\figwidth}
\begin{document}

\title{Avoiding irreversibility: engineering resonant conversions of quantum resources}

\author{Kamil Korzekwa}
\email{korzekwa.kamil@gmail.com}
\affiliation{Centre for Engineered Quantum Systems, School of Physics, University of Sydney, Sydney NSW 2006, Australia.}
\author{Christopher T.~Chubb}
\affiliation{Centre for Engineered Quantum Systems, School of Physics, University of Sydney, Sydney NSW 2006, Australia.}
\author{Marco Tomamichel}
\affiliation{Centre for Quantum Software and Information, School of Software, University of Technology Sydney, Sydney NSW 2007, Australia.}

\begin{abstract}
	We identify and explore the intriguing property of resource resonance arising within resource theories of entanglement, coherence and thermodynamics. While the theories considered are reversible asymptotically, the same is generally not true in realistic scenarios where the available resources are bounded. The finite-size effects responsible for this irreversibility could potentially prohibit small quantum information processors or thermal machines from achieving their full potential. Nevertheless, we show here that by carefully engineering the resource interconversion process any such losses can be greatly suppressed. Our results are predicted by higher order expansions of the trade-off between the rate of resource interconversion and the achieved fidelity, and are verified by exact numerical optimizations of appropriate approximate majorization conditions.
\end{abstract}

\maketitle

\paragraph*{Introduction.}

Rapid progress in experimental techniques to control intermediate-scale quantum systems~\cite{preskill2018quantum} may soon bring the advent of new technologies that will utilize quantum effects to overcome current limitations of electromechanical systems and information processors~\cite{dowling2003quantum}. From a theoretical perspective, the first step to achieve this is to identify which components of quantum theory can provide such an advantage, i.e., to recognize what actually constitutes quantum resources. Once identified, it is then crucial to understand when different resources can be interconverted, and how efficiently this can be done. The ultimate goal is to find \emph{reversible} resource transformations, i.e., transformations that can be perfectly reversed and that do not dissipate any resources. Such questions can be treated in great generality under the umbrella of quantum resource theories~\cite{coecke2016mathematical,chitambar2018quantum}.

With the above applications in mind, we focus on the question of reversibility of resource interconversion for intermediate-scale quantum systems, specifically in the resource theories of entanglement~\cite{horodecki2009quantum} and coherence~\cite{baumgratz2014quantifying} (for pure state transformations), and thermodynamics~\cite{horodecki2013fundamental} (for energy-incoherent transformations). In particular, we ask under what precise circumstances we can engineer reversible, and thus dissipationless, transformations between a relatively small number of resource states. In the above resource theories reversibility always holds in the {asymptotic regime}, where one assumes access to an infinite source of resource states~\cite{brandao2015reversible}. This means that, in the appropriate asymptotic limit, we can reverse the conversion process perfectly, recovering exactly the resource states we started with. Such an assumption, however, is unjustified in many practically relevant and fundamentally interesting scenarios. On the one hand, quantum resources like entanglement and coherence will only be available in small amounts in the near future; on the other hand, in quantum thermodynamics we want to explore how the known macroscopic laws change when we go beyond the thermodynamic limit and consider thermal processes involving a finite number of particles~\cite{goold2016role}.

A first attempt to address reversibility beyond the asymptotic limit was made in~\cite{kumagai2013entanglement,ito2015asymptotic} for the resource theory of entanglement. In~\cite{kumagai2017second} the authors developed a mathematical framework that allows us to study how irreversibility arises when we refine the asymptotic scenario and look at corrections that arise due to finite-size effects. This framework has recently been extended to the resource theory of thermodynamics by the present authors~\cite{chubb2017beyond}. On a technical level both these prior works deal with conversion processes that asymptotically recover the resource states only up to a constant error. The accompanying work~\cite{chubb18moderate} resolves this issue and considers processes that have asymptotically vanishing error. 

These analytical results together with the numerical results presented here reveal the phenomenon of \emph{resource resonance}, a surprising feature of resource interconversion that ensures that resource dissipation can be avoided by carefully engineering small quantum systems. Exploring this phenomenon and discussing its potential impact on quantum information processing and thermodynamic protocols will be the main focus of this paper.

\paragraph*{Motivating example.}

\begin{figure}[t]
	\begin{tikzpicture}
		\node (myfirstpic) at (0,0) {\includegraphics[width=\columnwidth]{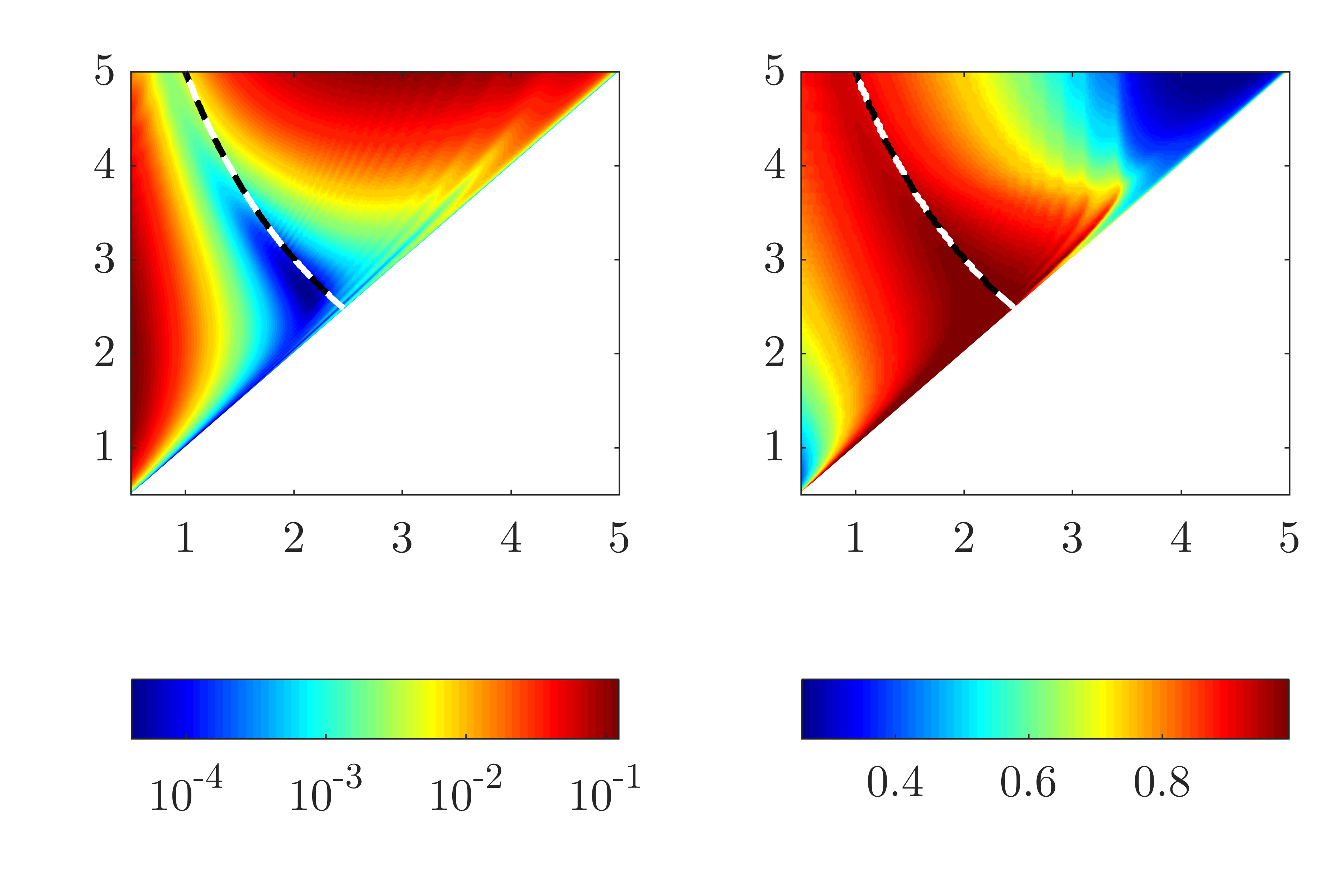}};		
		\node at (-0.475\columnwidth,0.315\columnwidth) {\small\color{black} (a)};
		\node at (0.02\columnwidth,0.315\columnwidth) {\small\color{black} (b)};
		\node at (0.285\columnwidth,-0.305\columnwidth) {\footnotesize\color{black} $\frac{W}{W_C}$};		
		\node at (-0.22\columnwidth,-0.305\columnwidth) {\footnotesize\color{black} $\epsilon$};		
		\node at (0.285\columnwidth,-0.125\columnwidth) {\footnotesize\color{black} $\frac{\Delta E}{k_B T_{c'}}$};		
		\node at (-0.22\columnwidth,-0.125\columnwidth) {\footnotesize\color{black} $\frac{\Delta E}{k_B T_{c'}}$};		
		\node[label={[label distance=0.5cm,text depth=-1ex,rotate=90]right:\footnotesize$\frac{\Delta E}{k_B T_c}$}] at (-0.51\columnwidth,0.015\columnwidth){};
		\node[label={[label distance=0.5cm,text depth=-1ex,rotate=90]right:\footnotesize$\frac{\Delta E}{k_B T_c}$}] at (-0.004\columnwidth,0.015\columnwidth){};		
	\end{tikzpicture}
	\caption{\label{fig:resonance_thermo} \emph{Resonance in work extraction.} Performance of the heat engine with a working body consisting of $n=200$ non-interacting qubits, each with energy gap $\Delta E$. Hot bath temperature it set by $k_BT_h=10\Delta E$ (where $k_B$ is the Boltzmann constant), the working body is initially at cold bath temperature $T_c$ and heats up to $T_{c'}$ in the process. (a) The optimal quality of work, measured by the infidelity $\epsilon$ between the final and excited battery state, while extracting $W=0.95W_C$ per qubit. (b) The optimal fraction $W_C$ that can be extracted per qubit when the quality of work is bounded by $\epsilon<10^{-3}$. The dashed line in both plots indicates the position of the resonance predicted by Eq.~\eqref{eq:irreversibility2}.}
\end{figure}
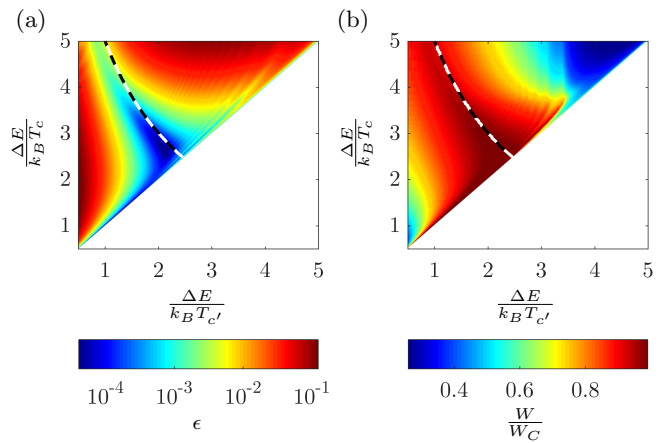

Before we delve into the details and present our results in full generality, let us first illustrate their spirit using the following simple example. Consider a heat engine, with a finite-size working body consisting of $n=200$ non-interacting two-level systems and operating between two thermal baths at temperatures $T_h>T_c$, performing work on a battery system initially in the ground state. As we explain later in the text, such a process can be elegantly phrased as an interconversion problem within the resource theory of thermodynamics, with the working body at cold temperature $T_c$ acting as a non-equilibrium resource in the presence of the hot bath at temperature~$T_h$. While the working body heats up from $T_c$ to $T_{c'}$, part of its resource content can be converted into work by exciting the battery system. Now, a perfect engine with an infinite working body, $n\rightarrow\infty$, and constantly operating at the Carnot efficiency would extract the amount of work $nW_C$ equal to the free energy change of the working body. However, since $n$ is small, the energy fluctuations of the working body are not negligible compared to the average energy, and thus the engine operates beyond the thermodynamic limit. We can then expect two kinds of effects. On the one hand, the quality of work~\cite{aberg2013truly,ng2017surpassing} will not be perfect, as some of the energy fluctuations will be transferred to the battery. In Fig.~\hyperref[fig:resonance_thermo]{1a} we present the optimal work quality as a function of $T_c$ and $T_{c'}$ for an engine extracting $W=0.95W_C$ per qubit of the working body. On the other hand, if we demand the work quality to be above some threshold level, the optimal efficiency of the engine may be affected so that it cannot achieve the Carnot limit. In Fig.~\hyperref[fig:resonance_thermo]{1b} we plot the optimal fraction of $W_C$ that can be extracted when its quality is bounded by a constant, again as a function of $T_c$ and $T_{c'}$. 

In both plots we see clear resonant lines, indicating near-perfect quality and near-Carnot efficiency corresponding to reversible (and thus dissipationless) processes, obtained while operating well outside the asymptotic regime with a finite-size working body. As we argue in this paper, it is in fact a genuine feature of all resource theories whose single-shot transformation laws are based on majorization relation. These include the considered resource theories of entanglement, coherence and thermodynamics, but also applies to, e.g., the studies of randomness conversions~\cite{kumagai2017second}. In all these cases, we find that if a pair of states satisfies a particular \emph{resonance condition}, one can achieve interconversion with dramatically reduced losses, i.e., transformation that is arbitrarily close to reversible even for finite number $n$ of resource states being processed. In what follows we first introduce necessary concepts, then state the resonance condition, and finally discuss how it can be employed to avoid irreversibility.

\paragraph*{Setting the scene.} 

The interconversion and dissipation of general quantum resources can be suitably analyzed within a resource-theoretic framework~\cite{coecke2016mathematical,chitambar2018quantum}. In a resource theory we allow only a subset of all possible quantum operations, and this in turn restricts the quantum states that can be prepared\,---\,the so-called free states of the theory. Natural restrictions may arise from practical difficulties, e.g., when preparing a system in a superposition of particular states is experimentally challenging, but may also be of fundamental nature, as with the laws of thermodynamics constraining possible transformations to preserve energy and increase entropy. 
A~quantum resource is any quantum system that, in conjunction with the allowed operations, allows one to overcome these restrictions. A paradigmatic example is provided by the resource theory of entanglement~\cite{horodecki2009quantum}: Alice, facing the restriction of not having access to a quantum channel to Bob, is unable to share an entangled quantum state $\ket{\psi}$ with him. However, if they are in possession of a maximally entangled state acting as a resource, they can use it to teleport Bob's share of~$\ket{\psi}$ using only local operations and classical communication. 

In this work we focus on the interconversion problem within the resource theories of entanglement, coherence and thermodynamics. These are defined via the relevant sets of free operations and free states, the latter of which we will collectively denote by $\omega$. More precisely, in entanglement theory one is restricted to local operations and classical communication (LOCC) and the allowed free states are given by separable states~\cite{horodecki2009quantum}; in coherence theory one is restricted to incoherent operations and the allowed free states are incoherent in the preferred basis~\cite{baumgratz2014quantifying}; finally, in the resource theory of thermodynamics (with respect to a fixed background temperature $T=1/\beta$) one is restricted to thermal operations and the only free state is a thermal equilibrium state~$\gamma$ at temperature $T$~\cite{janzing2000thermodynamic}. Rather surprisingly, all these three prominent examples of resource theories are formally very strongly related, as single-shot interconversion in each of them is ruled by a majorization partial order~\cite{nielsen1999conditions,du2015conditions} (or its variant known as thermo-majorization~\cite{horodecki2013fundamental}). This not only allows for a unified treatment, but also for a simplified representation of initial and target quantum states, $\rho$ and $\sigma$, as probability distributions, $\v{p}$ and $\v{q}$.

Before we describe this mapping, let us first explain the notions of multi-copy and approximate transformations. When considering transformations between multiple copies of initial and target states, $\rho^{\otimes n}$ and $\sigma^{\otimes Rn}$, the number of initial and target states will generally not be the same, i.e., conversion rate $R$ may differ from 1. However, since one can always append any number of free states $\omega$ to both the initial and target states, without loss of generality one can focus on transformations between total initial and target states,
\begin{equation}
	\tau_i=\rho^{\otimes n}\otimes\omega^{\otimes Rn},\quad \tau_f=\sigma^{\otimes Rn}\otimes\omega^{\otimes n}.
\end{equation}
Now, it may happen that for given $\tau_i$ and $\tau_f$ the interconversion is impossible, but there exists a state $\tilde{\tau}_f$ that is $\epsilon$-close to $\tau_f$ and such that the interconversion between $\tau_i$ and $\tilde{\tau}_f$ is possible. We then say that an approximate transformation is possible with the error level $\epsilon$ quantified by the infidelity between target state $\tau_f$ and final state~$\tilde{\tau}_f$.

As mentioned above, the interconversion conditions for all considered resource theories can be expressed using theory-dependent representations of initial and target quantum states. For entanglement theory, given initial and target pure bipartite states, $\rho=\ketbra{\Psi}{\Psi}$ and $\sigma=\ketbra{\Phi}{\Phi}$, with the Schmidt decomposition given by \mbox{$\ket{\Psi}=\sum_i a_i \ket{\psi_i}\otimes\ket{\psi_i}$} and \mbox{$\ket{\Phi}=\sum_i b_i \ket{\phi_i}\otimes\ket{\phi_i}$}, we can represent them via probability distributions
\begin{equation}
	\label{eq:ent:representation}
	p_i=|a_i|^2,\quad q_i=|b_i|^2.
\end{equation}
For coherence theory with respect to a fixed basis $\{\ket{i}\}$ one can represent pure initial and target states, \mbox{$\rho=\ketbra{\psi}{\psi}$} and $\sigma=\ketbra{\phi}{\phi}$, using 
\begin{equation}
	\label{eq:coh:representation}
	p_i=|\braket{i}{\psi}|^2,\quad q_i=|\braket{i}{\phi}|^2.
\end{equation}
Finally, in the resource theory of thermodynamics, the initial and target energy-incoherent mixed states $\rho$ and $\sigma$ can be represented by
\begin{equation}
	\label{eq:thermo:representation}
	p_i=\matrixel{E_i}{\rho}{E_i},\quad q_i=\matrixel{E_i}{\sigma}{E_i},
\end{equation}
where $\{\ket{E_i}\}$ denotes the energy eigenbasis of the system.  We will denote distributions representing free states $\omega$ by $\v{f}$. In entanglement and coherence theories these are represented by sharp probability distributions $\v{s}$ with a single non-zero entry; whereas in the thermodynamic case $\v{f}$ is given by a thermal Gibbs distribution $\v{\gamma}$ with $\gamma_i\propto\exp(-\beta E_i)$.

\paragraph*{Reversibility condition.} 

The optimal conversion rate $R$ between $n$ copies of an initial state $\rho$, represented by~$\v{p}$, and $Rn$ copies of the target state $\sigma$, represented by~$\v{q}$, can now be expressed in a unified way for all considered resource theories using a general asymptotic expansion,
\begin{equation}
	R \simeq R_\infty(\v{p},\v{q}) - r(\v{p},\v{q},n,\epsilon).
\end{equation}
In the above, $R_\infty$ denotes the rate achieved in the asymptotic limit, whereas $r$ is the correction term, which depends on the number of initial states $n$ and the accepted error level $\epsilon$, and vanishes for $n\to\infty$. Expressions for asymptotic rates $R_\infty$ are well known for the studied resource theories, with \mbox{$R_\infty=H(\v{p})/H(\v{q})$} for coherence and entanglement~\cite{bennett1996concentrating}, and \mbox{$R_\infty=D(\v{p}||\v{\gamma})/D(\v{q}||\v{\gamma})$} for thermodynamics~\cite{brandao2013resource}, where $H$ and $D$ denote the Shannon entropy and the relative entropy, respectively. What is important for the current analysis, is the fact that $R_\infty(\v{p},\v{q})=1/R_\infty(\v{q},\v{p})$, i.e., the asymptotic rate of a given process is equal to the inverse of the asymptotic rate for the reverse process. As a result, every resource interconversion is reversible in the asymptotic limit, so that no dissipation of resources takes place.

In the accompanying paper~\cite{chubb18moderate} we find the next order in the asymptotic expansion of the correction term $r$ for all considered resource theories in the \emph{moderate deviation} regime, i.e., when one demands both error and correction term to vanish with growing $n$~\footnote{Complementary to these results, are expressions for $r$ in the \emph{small deviation} regime (when one allows for constant error $\epsilon$ and vanishing $r$), obtained in \cite{kumagai2017second} and our previous work~\cite{chubb2017beyond}, for resource theories of entanglement and thermodynamics, respectively.}. The crucial observation is that for all $\alpha$ the correction term $r$ depends on a parameter \mbox{$\nu=\nu(\v{p},\v{q})$}, which we will refer to as the \emph{irreversibility parameter}. Its importance lies in the fact that as $\nu\geq 0$ gets closer to 1, the magnitude of $r$ diminishes, and in the limit $\nu=1$ it vanishes. Therefore, in the moderate deviation approximation, states satisfying $\nu=1$ can be interconverted with vanishing error at the asymptotic rate, $R=R_\infty$, even for finite $n$, and thus are reversibly interconvertible.

For entanglement and coherence transformations $\nu$ is given by~\cite{kumagai2017second}
\begin{equation}
	\label{eq:irreversibility1}
	\nu=\frac{V(\v{p})/H(\v{p})}{V(\v{q})/H(\v{q})},\quad	
\end{equation}
with $V$ denoting entropy variance~\cite{strassen62,hayashi08},
\begin{equation}
	V(\v{p})=\sum_i p_i \left( \log p_i + H(\v{p})\right)^2.
\end{equation}
Note that since the Shannon entropy $H$ quantifies the average (asymptotic) resource content of a state, entropy variance $V$ can be understood as quantifying its fluctuations. Indeed, $V$ vanishes only for distributions that are uniform on their support, which correspond to maximally coherent and maximally entangled states. Hence, the ratio $V/H$ tells us about relative strength of resource fluctuations in a single copy of a given state, and so the resonance condition is satisfied for states with equal relative resource fluctuations.

Similarly, in the thermodynamic case we have~\cite{chubb2017beyond}
\begin{equation}
\label{eq:irreversibility2}
	\nu=\frac{V(\v{p}||\v{\gamma})/D(\v{p}||\v{\gamma})}{V(\v{q}||\v{\gamma})/D(\v{q}||\v{\gamma})},
\end{equation}
	with $V$ denoting relative entropy variance~\cite{strassen62},
\begin{equation}
	V(\v{p}||\v{\gamma})=\sum_i p_i \left( \log \frac{p_i}{\gamma_i} - D(\v{p}||\v{\gamma})\right)^2.
\end{equation}
Note that $D$, quantifying resource content in the asymptotic (thermodynamic) limit, can be interpreted as non-equilibrium free energy of a state~\cite{brandao2013resource}. Thus, the relative entropy variance describes fluctuations of free energy and, again, states with equal relative fluctuations are in resonance. Moreover, as noted in~\cite{chubb2017beyond}, $V(\v{\gamma'}||\v{\gamma})$ (for $\v{\gamma}'$ being a thermal state at temperature $T'\neq T$) is proportional to the heat capacity at $T'$ and to the squared Carnot factor for $T$ and $T'$.

\paragraph*{Resource resonance.} 

The moderate deviation analysis of the asymptotic expansion for the resource conversion rate predicts a resonance phenomenon. However, the question that remains open is: how well does this analytical result capture the actual errors and rates arising in resource interconversion of a small number $n$ of resource states involved? To address this problem, we numerically investigate the interconversion problem for small-dimensional systems, and compare the results with our analytical predictions. More precisely, building on an algorithm developed by the present authors in~\cite{chubb2017beyond}, we find the actual optimal conversion rates for a fixed error level, and also the actual error level for a fixed conversion rate. The details can be found in the Supplementary Material, and here we focus on presenting a few illustrative examples.

\begin{figure}[t]
	\centering
	\setlength\figheight{4.1cm}
	\setlength\figwidth{4.1cm}
	\def\figscale{0.8}
\definecolor{mycolor1}{rgb}{0.00000,0.44700,0.74100}%
\definecolor{mycolor2}{rgb}{0.85000,0.32500,0.09800}%
\definecolor{mycolor3}{rgb}{0.92900,0.69400,0.12500}%
\definecolor{mycolor4}{rgb}{0.49400,0.18400,0.55600}%
\definecolor{mycolor5}{rgb}{0.46600,0.67400,0.18800}%
\definecolor{mycolor6}{rgb}{0.30100,0.74500,0.93300}%
\begin{tikzpicture}[%
scale=\figscale
]

\begin{axis}[%
width=0.951\figwidth,
height=\figheight,
at={(0\figwidth,0\figheight)},
scale only axis,
xmin=0.88,
xmax=1.15,
xlabel style={font=\color{white!15!black}},
xlabel={$\nu$},
ymode=log,
ymin=5e-06,
ymax=0.003,
yminorticks=true,
ylabel style={font=\color{white!15!black}},
ylabel={$\epsilon$},
ylabel near ticks,
axis background/.style={fill=white},
axis x line*=bottom,
axis y line*=left
]
\addplot [color=mycolor1, mark=*, mark options={solid, mycolor1}, forget plot]
  table[row sep=crcr]{%
1.14354423527767	0.00219352661573902\\
1.09166095747506	0.0017730572061927\\
1.03977769252716	0.0010038216215108\\
0.98789444043396	0.000575173301839205\\
0.936011201195458	0.000893198884463042\\
0.884127974811649	0.00144335706353815\\
};
\addplot [color=mycolor2, mark=*, mark options={solid, mycolor2}, forget plot]
  table[row sep=crcr]{%
1.14354423527767	0.00170844279454563\\
1.11760259476953	0.00144199721295257\\
1.09166095747506	0.00103141860078748\\
1.06571932339427	0.00071744597766199\\
1.03977769252716	0.000465761726300196\\
1.01383606487372	0.00025598851969677\\
0.98789444043396	0.000167296978933029\\
0.961952819207872	0.000261677439921115\\
0.936011201195458	0.000510083080993762\\
0.910069586396717	0.00085459745452332\\
0.884127974811649	0.00124284794424889\\
};
\addplot [color=mycolor3, mark=*, mark options={solid, mycolor3}, forget plot]
  table[row sep=crcr]{%
1.14354423527767	0.00160616418619963\\
1.12624980791517	0.00123130244989245\\
1.10895538198096	0.00106537393839123\\
1.09166095747506	0.000816925216470699\\
1.07436653439746	0.00060865714366809\\
1.05707211274816	0.000453026780721788\\
1.03977769252716	0.000298446593399504\\
1.02248327373446	0.000183879561442946\\
1.00518885637006	9.89156815391912e-05\\
0.98789444043396	7.71888647476127e-05\\
0.970600025926159	0.000130046540032458\\
0.953305612846659	0.000238512568234595\\
0.936011201195458	0.000386234566966626\\
0.918716790972556	0.000595039689947674\\
0.901422382177953	0.000838850476653663\\
0.884127974811649	0.00114209604044713\\
};
\addplot [color=mycolor4, mark=*, mark options={solid, mycolor4}, forget plot]
  table[row sep=crcr]{%
1.14354423527767	0.00137531017109438\\
1.13057341462189	0.00126390332455262\\
1.11760259476953	0.00100741154468076\\
1.10463177572059	0.000893973044776009\\
1.09166095747506	0.000715501212717684\\
1.07869014003296	0.000574240236260848\\
1.06571932339427	0.000448818555038866\\
1.05274850755901	0.000317587286884913\\
1.03977769252716	0.000223434195720351\\
1.02680687829873	0.000144129673973703\\
1.01383606487372	8.47319751834386e-05\\
1.00086525225213	4.3269274274782e-05\\
0.98789444043396	4.15474077856137e-05\\
0.974923629419207	7.55248594034308e-05\\
0.961952819207872	0.000137094048895281\\
0.948982009799956	0.000221815146997151\\
0.936011201195458	0.000324260619487315\\
0.923040393394378	0.000465978347317542\\
0.910069586396717	0.000638097984075325\\
0.897098780202474	0.000840120260074628\\
0.884127974811649	0.00108428513119763\\
};
\addplot [color=mycolor5, mark=*, mark options={solid, mycolor5}, forget plot]
  table[row sep=crcr]{%
1.14354423527767	0.00138507559626766\\
1.13316757868877	0.00116976928735502\\
1.12279092261406	0.00107425062796418\\
1.11241426705354	0.000891056096360754\\
1.10203761200721	0.000784633206444618\\
1.09166095747506	0.000646860965062168\\
1.08128430345711	0.000539976674693965\\
1.07090764995334	0.000441574844703529\\
1.06053099696376	0.000344324378072636\\
1.05015434448836	0.000264246925913314\\
1.03977769252716	0.000190077876805805\\
1.02940104108014	0.000122962487161704\\
1.01902439014732	7.37657428605543e-05\\
1.00864773972868	3.87220253684761e-05\\
0.998271089824224	2.10619462518435e-05\\
0.98789444043396	2.45538472837259e-05\\
0.977517791557884	5.06431564818444e-05\\
0.967141143195995	9.08656833474852e-05\\
0.956764495348295	0.00014149354444204\\
0.946387848014783	0.000208142784170029\\
0.936011201195458	0.000293855760336359\\
0.925634554890321	0.00038988877200985\\
0.915257909099371	0.000520154963736252\\
0.90488126382261	0.000672089307977131\\
0.894504619060036	0.000840024969891329\\
0.884127974811649	0.00105399005568529\\
};
\addplot [color=mycolor6, mark=*, mark options={solid, mycolor6}, forget plot]
  table[row sep=crcr]{%
1.14354423527767	0.00130622269044112\\
1.13489702141788	0.00119511499511793\\
1.12624980791517	0.00103254953582177\\
1.11760259476953	0.000942494798506055\\
1.10895538198096	0.000801089326192383\\
1.10030816954948	0.000718878739202844\\
1.09166095747506	0.000611614018302675\\
1.08301374575772	0.000518496474165464\\
1.07436653439746	0.000433034334907156\\
1.06571932339427	0.000351898932493988\\
1.05707211274816	0.000282937128037397\\
1.04842490245912	0.000218659960797951\\
1.03977769252716	0.000164449193532423\\
1.03113048295227	0.000118628389551456\\
1.02248327373446	7.46278025157698e-05\\
1.01383606487372	4.30742383471516e-05\\
1.00518885637006	1.94927512345844e-05\\
0.996541648223472	1.01509645568099e-05\\
0.98789444043396	1.81733516517735e-05\\
0.979247233001522	3.71828210133929e-05\\
0.970600025926159	6.51489993660492e-05\\
0.961952819207872	0.000102054040229338\\
0.953305612846659	0.000145386013952198\\
0.944658406842521	0.000202335248311702\\
0.936011201195458	0.000268274082563802\\
0.927363995905469	0.000344548580771264\\
0.918716790972556	0.000442899426564947\\
0.910069586396717	0.000556898658090388\\
0.901422382177953	0.000687619477580448\\
0.892775178316264	0.000846973875849089\\
0.884127974811649	0.0010205465394274\\
};
\addplot [color=black, dashed, forget plot]
  table[row sep=crcr]{%
1	5e-07\\
1	0.01\\
};
\end{axis}

\node at (-0.6,4.8) {\small\color{black} (a)};

\begin{axis}[%
width=0.951\figwidth,
height=\figheight,
at={(0\figwidth,0\figheight)},
scale only axis,
xmin=0.88,
xmax=1.15,
xtick={0.884,0.949,1.014,1.079,1.144},
xticklabels={{0},{0.25},{0.5},{0.75},{1}},
xlabel style={at={(axis description cs:0.5,1.35)},font=\color{white!15!black}},
xlabel={$\lambda$},
ymode=log,
ymin=0.1,
ymax=1,
ytick={\empty},
yminorticks=true,
axis x line*=top,
axis y line*=right
]
\end{axis}
\end{tikzpicture}%
	\hspace{0.1cm}
\definecolor{mycolor1}{rgb}{0.00000,0.44700,0.74100}%
\definecolor{mycolor2}{rgb}{0.85000,0.32500,0.09800}%
\definecolor{mycolor3}{rgb}{0.92900,0.69400,0.12500}%
\definecolor{mycolor4}{rgb}{0.49400,0.18400,0.55600}%
\definecolor{mycolor5}{rgb}{0.46600,0.67400,0.18800}%
\definecolor{mycolor6}{rgb}{0.30100,0.74500,0.93300}%
\begin{tikzpicture}[%
scale=\figscale
]

\begin{axis}[%
width=0.951\figwidth,
height=\figheight,
at={(0\figwidth,0\figheight)},
scale only axis,
xmin=4.5,
xmax=30.5,
xlabel style={font=\color{white!15!black}},
xlabel={$n$},
ymin=0,
ymax=0.85,
ylabel style={font=\color{white!15!black}},
ylabel={$R$},
ylabel near ticks,
axis background/.style={fill=white},
axis x line*=bottom,
axis y line*=left
]
\addplot [color=red, draw=none, mark size=2.0pt, mark=*, mark options={solid, mycolor1}, forget plot]
  table[row sep=crcr]{%
5	0.6\\
6	0.5\\
7	0.571428571428571\\
8	0.5\\
9	0.555555555555556\\
10	0.6\\
11	0.545454545454545\\
12	0.583333333333333\\
13	0.615384615384615\\
14	0.571428571428571\\
15	0.6\\
16	0.625\\
17	0.588235294117647\\
18	0.611111111111111\\
19	0.631578947368421\\
20	0.6\\
21	0.619047619047619\\
22	0.636363636363636\\
23	0.608695652173913\\
24	0.625\\
25	0.64\\
26	0.615384615384615\\
27	0.62962962962963\\
28	0.607142857142857\\
29	0.620689655172414\\
30	0.633333333333333\\
};
\addplot [color=black, dotted, line width=2.0pt, forget plot]
  table[row sep=crcr]{%
0	0.799999938052323\\
100	0.799999938052323\\
};
\addplot [color=green, draw=none, mark size=2.0pt, mark=*, mark options={solid, mycolor5}, forget plot]
  table[row sep=crcr]{%
5	0.2\\
6	0.333333333333333\\
7	0.285714285714286\\
8	0.375\\
9	0.333333333333333\\
10	0.4\\
11	0.363636363636364\\
12	0.416666666666667\\
13	0.384615384615385\\
14	0.428571428571429\\
15	0.4\\
16	0.4375\\
17	0.411764705882353\\
18	0.444444444444444\\
19	0.473684210526316\\
20	0.45\\
21	0.476190476190476\\
22	0.454545454545455\\
23	0.478260869565217\\
24	0.5\\
25	0.48\\
26	0.5\\
27	0.481481481481481\\
28	0.5\\
29	0.517241379310345\\
30	0.5\\
};
\addplot [color=black, dotted, line width=2.0pt, forget plot]
  table[row sep=crcr]{%
0	0.799999879200486\\
100	0.799999879200486\\
};
\addplot [color=blue, draw=none, mark size=2.0pt, mark=*, mark options={solid, mycolor2}, forget plot]
  table[row sep=crcr]{%
5	0\\
6	0\\
7	0\\
8	0.125\\
9	0.111111111111111\\
10	0.2\\
11	0.181818181818182\\
12	0.25\\
13	0.230769230769231\\
14	0.214285714285714\\
15	0.266666666666667\\
16	0.3125\\
17	0.294117647058824\\
18	0.333333333333333\\
19	0.315789473684211\\
20	0.3\\
21	0.333333333333333\\
22	0.363636363636364\\
23	0.347826086956522\\
24	0.375\\
25	0.36\\
26	0.384615384615385\\
27	0.407407407407407\\
28	0.392857142857143\\
29	0.413793103448276\\
30	0.4\\
};
\addplot [color=black, dotted, line width=2.0pt, forget plot]
  table[row sep=crcr]{%
0	0.799999936724604\\
100	0.799999936724604\\
};
\end{axis}

\node at (-0.6,4.8) {\small\color{black} (b)};

\node at (2.7,3.4) {\scriptsize\color{mycolor1}$\nu=1.95$};
\node at (2.7,2.65) {\scriptsize\color{mycolor5}$\nu=3.26$};
\node at (2.7,1.2) {\scriptsize\color{mycolor2}$\nu=4.25$};

\begin{axis}[%
width=0.951\figwidth,
height=\figheight,
at={(0\figwidth,0\figheight)},
scale only axis,
xmin=4.5,
xmax=30.5,
xtick={\empty},
ymin=0,
ymax=0.85,
ytick={\empty},
axis x line*=top,
axis y line*=right
]
\end{axis}

\end{tikzpicture}%
	\caption{\label{fig:resonance_ent} \emph{Tuning resources to resonance.} State interconversion under LOCC for a bipartite system consisting of $n$ pairs of qutrits. (a) The infidelity $\epsilon$ between the target state $\ket{\Phi}^{\otimes n}$ and the optimal final state obtained from \mbox{$\ket{\Psi_1}^{\otimes \lambda n}\otimes\ket{\Psi_2}^{\otimes (1-\lambda)n}$}. Different plots correspond to varying numbers of interconverted states \mbox{$n\in\{5,10,\dots,30\}$}, from top to bottom. The location of the resonant mixing factor $\lambda_*$ (dashed line) can be found using Eq.~\eqref{eq:irreversibility1}. (b)~The optimal conversion rate $R$ between the initial state $\ket{\Psi_i}$ and the target state $\ket{\Phi}^{\otimes Rn}$ with transformation infidelity bounded by $\epsilon<0.01$. Different plots correspond to initial states with equal asymptotic conversion rate $R_\infty$ (dashed line), but varying irreversibility parameters~$\nu$. The Schmidt coefficients of all states can be found in the Supplementary Material.}
\end{figure}
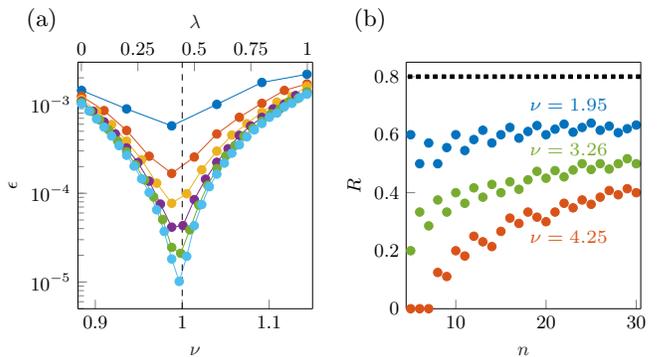

We start by considering the following scenario: imagine one has access to $n$ systems, each initially in a state~$\rho_1$ or $\rho_2$, and wants to transform them to a target state~$\sigma$. Moreover, assume that the asymptotic resource values of $\rho_1$ and $\rho_2$ are equal, i.e., asymptotically one can obtain the same number of copies $R_\infty n$ of $\sigma$ from either $n$ copies of $\rho_1$ or $n$ copies of $\rho_2$. One can achieve this asymptotic conversion rate also for finite $n$, but for the price of error~$\epsilon$. This error depends on the irreversibility parameter~$\nu$, and thus for finite $n$ the states $\rho_1$ and $\rho_2$ no longer have the same value. Crucially, however, it may happen that they are incompatible with $\sigma$ in opposite ways, such that the irreversibility parameter for $\rho_1$ and~$\sigma$ is smaller than~$1$, while for $\rho_2$ and $\sigma$ it is larger than 1. Thus, by taking $\lambda n$ copies of $\rho_1$ and $(1-\lambda)n$ copies of $\rho_2$ one can tune the initial state to be in resonance with the target state, i.e., to have irreversibility parameter close to 1. In Fig.~\hyperref[fig:resonance_ent]{2a} we show how such tuning can reduce the infidelity of entanglement transformation by several orders of magnitude. Note that this effect is much stronger than the expected increase in fidelity due to the increased number $n$ of processed states that can be observed for $\lambda \in \{0, 1\}$.

We also consider an alternative situation where, instead of demanding conversion at the asymptotic rate, we enforce the error to be below some fixed threshold value $\epsilon$. For the sake of simplicity, we can again focus on a set of initial states $\{\rho_i\}$ that are asymptotically equivalent, and ask how many copies of the target state $\sigma$ can we optimally get for $\rho_i^{\otimes n}$, with error not exceeding $\epsilon$. Now, based on our analytical results, we expect that as $n$ grows the conversion rate $R$ will approach the asymptotic value $R_\infty$ quicker for states $\rho_i$ that are closer to resonance with $\sigma$. Indeed, this is the case, as the numerical results presented in Fig.~\hyperref[fig:resonance_ent]{2b} show. 

Finally, let us come back to our initial motivating example (see Fig.~\ref{fig:resonance_thermo}) to explain it in more detail. First of all, we note that it is a particular instance of the paradigmatic thermodynamic protocol of work extraction~\cite{horodecki2013fundamental,aberg2013truly,skrzypczyk2014work}, where the state we extract work from is given by a thermal state at a colder temperature than the background temperature $T_h$. More precisely, our aim is to transform $n$ copies of $\gamma_c\otimes\ketbra{0}{0}$ (where the first system refers to the working body and second to the battery consisting of $n$ qubits) into $n$ copies of $\gamma_{c'}\otimes\ketbra{1}{1}$, with subscripts $c$ and $c'$ denoting initial and final temperatures of the working body, $T_c$ and $T_{c'}$, and $\ketbra{1}{1}$ being the excited state of the battery with energy $W$. The amount of extracted work is then given by $n W$ and its quality is measured by the infidelity between the battery's final state and its target state $\ketbra{1}{1}^{\otimes n}$. It is known that in the asymptotic limit the optimal thermodynamic process extracts $nW_C$ with \mbox{$W_C=k_BT_h[D(\v{\gamma}_c||\v{\gamma}_h)-D(\v{\gamma}_{c'}||\v{\gamma}_h)]$}~\cite{brandao2013resource}, which coincides with the amount of work that would be extracted by an engine operating at Carnot efficiency between an infinite bath at fixed temperature $T_h$ and a finite working body that heats up during the process from $T_c$ to $T_{c'}$~\cite{chubb2017beyond}. For small $n$, however, fluctuations can decrease this optimal amount (alternatively, decrease the quality of extracted work), as already discussed before and presented in Fig.~\ref{fig:resonance_thermo}. It is also important to note that, since $V(\v{\gamma}_c||\v{\gamma}_h)$ is positive and vanishes for $T_c=T_h$ and $T_c=0$, for all $T_h$ there exist resonant pairs of temperatures $T_c$ and $T_{c'}$, for which fluctuations vanish.

\paragraph*{Conclusions.}

We have shown that finite-size analysis can bring qualitatively new insights into the nature of interconversion processes within resource theories of entanglement, coherence and thermodynamics. 
This is in rather stark contrast to the small or moderate deviation analysis for information theoretic tasks based on quantum hypothesis testing, including many channel coding problems (see, e.g.,~\cite{tomamichel12,li12,tomamicheltan14,chubb17,cheng17}), where we instead see a behaviour that can be understood as a rather immediate manifestation of the central limit theorem and its moderate deviation analogue. Here, we have demonstrated that the predicted \emph{finite resource resonance} is clearly visible numerically and is not dominated by higher order asymptotics neglected in the analytical approximations. This makes us believe that the resonance effect could also be observed in noisy intermediate-scale quantum devices in the near future. Moreover, we explained how the predicted phenomenon can be employed to significantly improve the quality of the conversion process. 

On the one hand, our result may serve as a guiding principle for devising optimal protocols for resource conversion that would minimize losses by operating near resonance. On the other hand, it strongly motivates the investigation of the second order asymptotic corrections to interconversion rates in other resource theories. In particular, one could look for such corrections within the resource theory of $U(1)$ asymmetry~\cite{marvianthesis,marvian2013asymmetry}, which could contribute to our understanding of quantum thermodynamics beyond energy-incoherent states~\cite{lostaglio2015quantum,cwiklinski2015limitations}. Finally, while it is true that the resonance phenomenon is strongly related to the majorization condition, it is possible that similar behaviour can be observed in other quantum information processing tasks~\cite{tomamichel2015quantum}.

\paragraph{Acknowledgements.}

We thank Masahito Hayashi for helpful comments. We acknowledge support from the ARC via the Centre of Excellence in Engineered Quantum Systems (EQUS), project number CE110001013. CC also acknowledges support from the Australian Institute for Nanoscale Science and Technology Postgraduate Scholarship (John Makepeace Bennett Gift). MT acknowledges an Australian Research Council Discovery Early Career Researcher Award, project no. DE160100821.

\bibliography{thermo,library_mt}

\clearpage
\setcounter{equation}{0}
\renewcommand{\theequation}{S\arabic{equation}}
\onecolumngrid
\widetext
\begin{center}
	\textbf{\large Supplementary Material}
\end{center}
\bigskip

Here we explain how the numerical data presented in the plots of the main text was obtained. First, note that the approximate interconversion conditions are ultimately specified by approximate majorization and thermo-majorization relations. In order to find the final distribution that is closest in infidelity distance to the target state, and at the same time satisfies the required majorization relation, we employ the algorithm developed in our previous work~\cite{chubb2017beyond} (and conceptually based on \cite{vidal2000approximate}). The precise details can be found in Appendix~C of~\cite{chubb2017beyond} (the \texttt{python} code is also attached to the arXiv submission of~\cite{chubb2017beyond}). 

The data presented in Figs.~\hyperref[fig:resonance_thermo]{1a}~and~\hyperref[fig:resonance_thermo]{1b} was obtained in the following way. First, the temperature and energy units were chosen such that the Boltzmann constant $k_B=1$, and the energy difference between the ground and excited state of the qubit $\Delta E=1$. Thus, thermal distributions at temperature $T_x$ of qubits comprising the working body are given by
\begin{equation}
	\v{\gamma}_x=\frac{1}{1+e^{-1/T_x}} \left[1,e^{-1/T_x}\right].
\end{equation}
The hot temperature was set to $T_h=10$, while the initial and final cold temperatures, $T_c$ and $T_{c'}$, varied between $0.5$ and $5$ (with $T_{c'}>T_c$). Now, for each pair of points $(T_c,T_{c'})$ the following free energy difference was calculated,
\begin{equation}
	W_C(T_c,T_{c'}):=T_h[D(\v{\gamma}_c||\v{\gamma}_h)-D(\v{\gamma}_{c'}||\v{\gamma}_h)].
\end{equation}
Note that this is the optimal amount of work that can be extracted in the asymptotic limit per one qubit of the working body. The process of extracting work $W$ per copy is modelled by the following interconversion process
\begin{equation}
	\label{supp:eq:conversion}
	(\v{\gamma_c}\otimes[1,0])^{\otimes n}\longrightarrow 	(\v{\gamma_{c'}}\otimes[0,1])^{\otimes n},
\end{equation}
where the second system is a qubit battery with the energy difference between the ground and excited state set to~$W$. For Fig.~\hyperref[fig:resonance_thermo]{1a} we set $W=0.95W_C$ and run the algorithm yielding the optimal infidelity $\epsilon$ between the right hand side of Eq.~\eqref{supp:eq:conversion} (the target state) and a distribution thermo-majorized by the left hand side of Eq.~\eqref{supp:eq:conversion} (the final state). For Fig.~\hyperref[fig:resonance_thermo]{1b} we set $W=x W_C$ with $x$ varying from $1$ to $0$, until the same algorithm does not yield infidelity $\epsilon$ below the threshold value set to $10^{-3}$. For both plots we set $n=200$.

We now proceed to Figs.~\hyperref[fig:resonance_ent]{2a}~and~\hyperref[fig:resonance_ent]{2b}. Pure state transformations under LOCC are fully determined by majorization relations between Schmidt coefficients of the transformed states. Let us then denote the Schmidt coefficients of a pure bipartite state $\ket{\Psi}$ by~$\v{\zeta}_{\Psi}$. The data presented in Fig.~\hyperref[fig:resonance_ent]{2a} was obtained for initial and target states characterized by the following Schmidt coefficients (up to 4 significant digits):
\begin{subequations}
\begin{align}
	\v{\zeta}_{\Psi_1}=&[0.4309,0.4300,0.1391],\quad H(\v{\zeta}_{\Psi_1})=1,\quad V(\v{\zeta}_{\Psi_1})=0.1529,\\
	\v{\zeta}_{\Psi_2}=&[0.5499,0.2300,0.2201],\quad H(\v{\zeta}_{\Psi_2})=1,\quad V(\v{\zeta}_{\Psi_2})=0.1977,\\
	\v{\zeta}_{\Phi_{\phantomsection{1}}}=&[0.5121,0.3300,0.1579],\quad H(\v{\zeta}_{\Phi})_{\phantom{1}}=1,\quad V(\v{\zeta}_{\Phi})_{\phantom{1}}=0.1729.
\end{align}
\end{subequations}
Similarly, Fig.~\hyperref[fig:resonance_ent]{2b} was obtained using:
\begin{subequations}
	\begin{align}
	\v{\zeta}_{\Psi_1}=&[0.5436,0.4264,0.0300],\quad H(\v{\zeta}_{\Psi_1})=0.8,\quad V(\v{\zeta}_{\Psi_1})=0.2406,\\
	\v{\zeta}_{\Psi_2}=&[0.6594,0.2806,0.0600],\quad H(\v{\zeta}_{\Psi_2})=0.8,\quad V(\v{\zeta}_{\Psi_2})=0.4024,\\
	\v{\zeta}_{\Psi_3}=&[0.7119,0.1481,0.1400],\quad H(\v{\zeta}_{\Psi_3})=0.8,\quad V(\v{\zeta}_{\Psi_3})=0.5236,\\	
	\v{\zeta}_{\Phi_{\phantom{1}}}=&[0.4514,0.4086,0.1400],\quad H(\v{\zeta}_{\Phi})_{\phantom{1}}=1,\phantom{.8}\quad V(\v{\zeta}_{\Phi})_{\phantom{1}}=0.1541.
	\end{align}
\end{subequations}
As in the thermodynamic case, here also we employed the algorithm yielding the optimal infidelity between the final and target states.

\end{document}